\documentclass{article}

\usepackage[nonatbib,final]{neurips_2024_ml4ps}

\usepackage[utf8]{inputenc} 
\usepackage[T1]{fontenc}    
\usepackage{hyperref}       
\usepackage{url}            
\usepackage{booktabs}       
\usepackage{amsfonts}       
\usepackage{nicefrac}       
\usepackage{microtype}      
\usepackage{xcolor}         

\usepackage{amsmath}
\usepackage{graphicx}
\usepackage{subfigure}

\usepackage{algorithm}
\usepackage{algorithmic}
\usepackage{wrapfig}

\title{Loss function to optimise signal significance in particle physics}

\author{%
  Jai Bardhan$^1$\\
  \texttt{jai.bardhan90@gmail.com}\\
  \And
  Cyrin Neeraj$^1$\\
  \texttt{cyrin.neeraj@research.iiit.ac.in}\\
  \And 
  Subhadip Mitra$^1$\\
  \texttt{subhadip.mitra@iiit.ac.in}\\
  \And
  Tanumoy Mandal$^2$\\
  \texttt{tanumoy.mandal@iisertvm.ac.in}\\
  \And
  $^1$\normalfont{Center for Computational Natural Science and Bioinformatics,} \\
  International Institute of Information Technology, Hyderabad
  \And
  $^2$\normalfont{Indian Institute of Science Education and Research, Thiruvananthapuram} 
}

\begin{document}

\maketitle

\begin{abstract}

  We construct a surrogate loss to directly optimise the significance metric used in particle physics. We evaluate our loss function for a simple event classification task using a linear model and show that it produces decision boundaries that change according to the cross sections of the processes involved. We find that the models trained with the new loss have higher signal efficiency for similar values of estimated signal significance compared to ones trained with a cross-entropy loss, showing promise to improve sensitivity of particle physics searches at colliders.
  
\end{abstract}

\section{Introduction}
Particle physics experiments at the Large Hadron Collider (LHC) rely heavily on multivariate classifiers to isolate signals from backgrounds. These investigations are generally of two types: 1) measuring known processes/properties with improving precision and checking for anomalies [i.e., departures from the predictions of the Standard Model] and 2) looking for new processes (like looking for hypothetical particles). In most cases, the need for multivariate classifiers comes from the sporadic nature of the signal compared to the background. 
Generally, the signal plus background hypothesis ($H_1$) is tested against the null or background-only hypothesis ($H_0$), and the disagreement between them is expressed in terms of a $p$ value. An equivalent interpretation of the $p$ value is the significance score ($Z$) defined such that a Gaussian-distributed variable found $Z$ standard deviations away from its mean has a tail-distribution probability equal to $p$~\cite{Cowan:2010js}. Most sensitivity studies commonly use a simple approximation of the median $Z$ score as a measure of the estimated signal significance,
\begin{equation}
    \label{eq:ZscoreApprox}
    Z \approx N_s/\sqrt{N_b}
\end{equation}
where $N_s$ and $N_b$ are the estimated numbers of signal and background events, respectively. (In the rest of the paper, we shall refer to the med[$Z$] score as just the $Z$ score.) 

In this paper, we attempt to construct a loss function whose minimisation can directly enhance the experimental sensitivity. Our motivation comes from two observations. First, not all event rates are equal; some scattering processes have higher probabilities [parameterised as cross sections ($\sigma$), calculable from theory. The number of events from a process ($N_{s,b}$) is calculated as $N_{s,b}=\sigma_{s,b}\mathcal L$, where $\mathcal L$ is the experimental luminosity] than others. Since a basic binary cross-entropy (BCE) loss treats all events equally, it might wrongly classify events of some types which is more detrimental to the classifier performance than the other. Second, a loss that optimises the signal-to-background ratio ($r=N_s/N_b$) may not necessarily maximise the significance as the $Z$ score depends on the ratio and the absolute number (i.e., set size) of signal or background events ($Z\approx \sqrt{N_s}\cdot\sqrt{r} = r\sqrt{N_b}$). Hence we ask, can we derive a loss function that maximises the $Z$ score directly? 
The $Z$ score described in Eq.~\eqref{eq:ZscoreApprox} is a set function. Therefore, we define a surrogate loss function using the Lov\'{a}sz extension~\cite{DBLP:journals/corr/BermanB17} to maximise it directly. We evaluate this loss with pseudo-data mimicking a typical event classification task using a linear model and compare the decision boundaries to that model trained on a BCE loss. We also compare the performance of models trained on a BCE loss.

\section{Constructing the loss: submodularity and Lov\'asz's extension}

We must consider some points before constructing a loss function based on the $Z$ score. First, since the $Z$ score is not a differentiable function (it depends upon discrete quantities), it needs a smooth interpolation. Second, the metric operates on datasets instead of individual samples---particularly, count data. Therefore, we must either develop a method to directly optimise the set function or assign contributions to specific samples within the set to optimise. 

We look for a smooth submodular function. A submodular function is a function that captures the concept of diminishing returns. It is defined on sets and has a property similar to concavity. Formally, submodularity can be defined as:

\paragraph{Submodularity:}
    A set function $\Delta: \{0, 1\}^p \to \mathbb{R}$ is submodular if for all sets $A, B \in \{0, 1\}^p$,
    \begin{equation}
        \Delta(A) + \Delta(B) \geq \Delta(A \cup B) + \Delta (A \cap B),
        \label{eq:somelab}
    \end{equation}
or, equivalently for $B \subseteq A$ and $i \notin A, i \notin B$, 
    \begin{equation}
        \Delta(A \cup \{i\}) - \Delta(A) \leq \Delta(B \cup \{i\}) - \Delta(B).
    \end{equation}

The submodular functions can be optimised using greedy optimisation techniques, and it is to find optimal solutions in polynomial times. However, these discrete optimisation techniques cannot be used directly without a gradient. The Lov\'{a}sz extension allows us to associate a continuous, convex function with any submodular function: 
\paragraph{Lov\'{a}sz extension:}
    For a set function $\Delta: \{0, 1\}^p \to \mathbb{R}$, the Lov\'{a}sz extension $\Bar{\Delta}: [0, 1]^p \to \mathbb{R}$ is defined as
    \begin{equation}
        \label{eq:lovaszExten}
        \Bar{\Delta}: \mathbf{m} \in \mathbb{R}^p \mapsto
        \sum_{i=1}^p m_i\ g_i(\mathbf{m})
    \end{equation}
where $\mathbf{m} \in \mathbb{R}^p_+$ is the vector of errors (which we discuss in the next section), $g_i(\mathbf{m}) = \Delta(\{\pi_1, \dots, \pi_i\}) - \Delta(\{\pi_i, \dots, \pi_{i-1}\})$ and $\boldsymbol{\pi}$ is a permutation ordering the components of $\mathbf{m}$ in decreasing order, i.e., $x_{\pi_1} \geq x_{\pi_2} \geq \dots \geq x_{\pi_p}$~\cite{DBLP:journals/corr/BermanB17}.
For the Lov\'{a}sz extension to be applicable, the set function must be submodular.

Additionally, the Lov\'asz extension of a submodular function preserves submodularity, i.e., the extension evaluated at the points of the hypercube still follows submodularity. Using the Lov\'{a}sz extension, we can directly compute the tight convex closure of a submodular function within polynomial time [$\mathcal{O}(p \log(p))$ time complexity]. This convex extension is amenable to a host of efficient optimisation methods, especially gradient-based approaches. 

\subsection{Submodularity of the $\mathbf{Z}$ score}
\label{sec:z_score}
 
We modify the $Z$ score defined in Eq.~\eqref{eq:ZscoreApprox} by setting $N_b = \epsilon + N_b$ to prevent the term from diverging at $N_b = 0$. We also add a minus sign to make it suitable for minimisation. The surrogate set function for the $Z$-score is then given by,
\begin{equation}
    \label{eq:LZ-delta-z}
    \Delta_Z(y, \Tilde{y},\boldsymbol{\sigma},\boldsymbol{v}, \mathcal{L}) = \sum_{i \in S} \frac{\sigma_i \mathcal{L}}{\sqrt{\epsilon}} - \frac{\sum_{i \in S} \frac{v_i - n_i}{v_i} \sigma_i \mathcal{L}}{\sqrt{ \epsilon + \sum_{i \in B} \frac{p_i}{v_i} \sigma_i \mathcal{L}}}
\end{equation}
where $y$ stands for the ground-truth labels of a set of events and $\Tilde{y}$ for the labels predicted by a model on the set; $P_y$ and $P_{\Tilde{y}}$ represent the set of positive labels and positive predictions and $v_i$ is the number of events of process type $i$ where $i\in S \cup B$, where $S, B$ are the set of signal and background processes. The constant term $\left(\sum_{i \in S} \sigma_i \mathcal{L}/\sqrt{\epsilon}\right)$ is added to ensure $\Delta_Z(\emptyset) = \boldsymbol{0}$.
The function $\Delta_Z$ is submodular on the set of mispredictions $(n,p)$, where $n$ is the number of false negatives, and $p$ is the number of false positives (which can be calculated from $P_y$ and $P_{\Tilde{y}}$). The proof is presented in Appendix~\ref{App:Z_submodular_proof}. Even though here we consider only one signal process, the proof can be trivially generalised to the cases with multiple signal processes ($n_1, n_2, \dots, n_r$) as well.
From Eq.~\eqref{eq:lovaszExten}, we see that for $\Delta_Z$ to be a loss function, the vector $\mathbf{m}$ must be the error vector in the prediction; $\Bar{\Delta}_Z$, then naturally emerges as the surrogate loss to optimise the $Z$ score. 


\paragraph{Choice of $\mathbf{m}$ error:}\label{sec:LZerr}
There are a few choices in the literature for modelling the error vector $m$. To illustrate the working of the $Z$-score loss for this paper, we pick a hinge (Max Margin) error similar to Ref.~\cite{lovaszHinge}. The labels are considered signed ($y_i \in \{-1, 1\}$). The model outputs a score $F_i(x)$ for each sample $x$. The error is given by the hinge loss, 
\begin{equation}
    m_i = \max(1 - F_i(x)y_i, 0), \qquad y_i \in \{-1, 1\}.
\end{equation}
The vector $\mathbf{m}$  could also be modelled as a sigmoid error or cross-entropy error, for example. We plot the $Z$-score loss landscape for all these errors in the appendix for the toy problem (described below) in Appendix~\ref{app:error_fns}.

There is only one free parameter in our loss: $\epsilon$. Other quantities like $\sigma_i$ and $\mathcal{L}$ are set by the process under consideration (i.e., the particular classification task) and the collider experiment. Assuming we perform the classification for rare signals, we set $\epsilon = \sigma_{s}\mathcal{L}$, the theoretically predicted number of signal events (which is also the maximum number of estimated signal events) for testing the loss. 
Algorithm~\ref{alg:lovaszGradCalc} provides a simple pseudocode to calculate the gradient $g(\mathbf{m})$ from Eq.~\eqref{eq:lovaszExten} using Eq.~\eqref{eq:LZ-delta-z} as the loss.

\begin{wrapfigure}[12]{R}{0.6\textwidth}
\vspace{-7.0em}
\begin{minipage}{0.6\textwidth}
    \begin{algorithm}[H]
       \caption{Gradient of Lov\'{a}sz $Z$ loss $\Bar{\Delta}_Z$}
       \label{alg:lovaszGradCalc}
    \begin{algorithmic}[1]
       \REQUIRE vector of errors $\mathbf{m} \in \mathbb{R}_+^p$, ground truth labels $\boldsymbol{\delta}$, sample weights $\boldsymbol{w} = \{w_1, w_2, \dots, w_p\}$ calculated from $\sigma_i$ and counts.
       \ENSURE $g(\mathbf{m})$ gradient of $\Bar{\Delta}_Z$ from Equation \eqref{eq:lovaszExten}
       \STATE {$\boldsymbol{\pi} \leftarrow$} decreasing sort permutation for $\boldsymbol{m}$
       \STATE {$\boldsymbol{\delta_\pi} \leftarrow \left(\delta_{\pi_i}\right)_{i \in [1,p]}$} 
       \STATE {\bfseries numerator $\leftarrow$} 1 - \textbf{cumulative\_sum$\boldsymbol{(\delta_\pi)}$} $\boldsymbol{w}$
       \STATE {\bfseries denominator $\leftarrow$} 1 + \textbf{cumulative\_sum$\boldsymbol{(1 - \delta_\pi)}$} $\boldsymbol{w}$
       \STATE {$\boldsymbol{g} \leftarrow$} $\sigma - \textbf{numerator}/{\sqrt{\textbf{denominator}}}$
       \IF{$p > 1$}
       \STATE $\boldsymbol{g}[2 : p] \leftarrow \boldsymbol{g}[2 : p] - \boldsymbol{g}[1 : p - 1]$
       \ENDIF
       \STATE {\bfseries return} $\boldsymbol{g_\pi}$
    \end{algorithmic}
    \end{algorithm}
\end{minipage}
\end{wrapfigure}

\section{Testing the loss}\label{sec:3}
\begin{figure*}[th]
\centering
\subfigure[\hspace{-0.75cm}]{\includegraphics[width=0.4\textwidth]{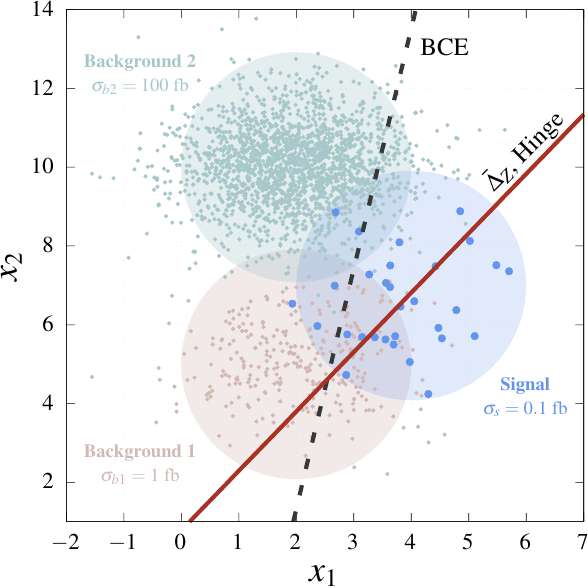}}\hspace{1cm}
\subfigure[\hspace{-0.75cm}]{\includegraphics[width=0.4\textwidth]{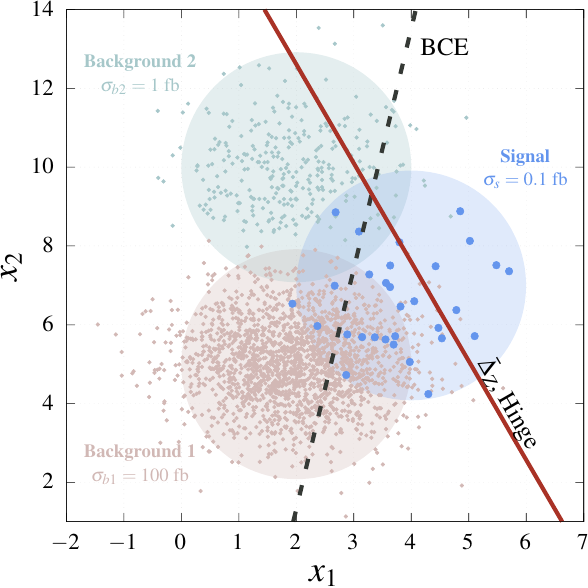}}
\caption{\label{fig:Loss_DecBound} Decision boundaries of the linear classifier when trained with the $\bar{\Delta}_Z$ loss with hinge error for (a) Case 1 and (b) Case 2 (see Section~\ref{sec:3}).  When trained with $\bar{\Delta}_Z$, the classifier prioritises reducing the background with the larger cross section.}
\subfigure[\hspace{-0.5cm}]{\includegraphics[width=0.31\textwidth]{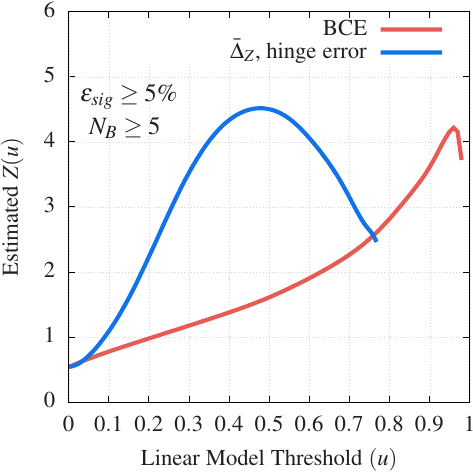}}\hfill
\subfigure[\hspace{-0.5cm}]{\includegraphics[width=0.31\textwidth]{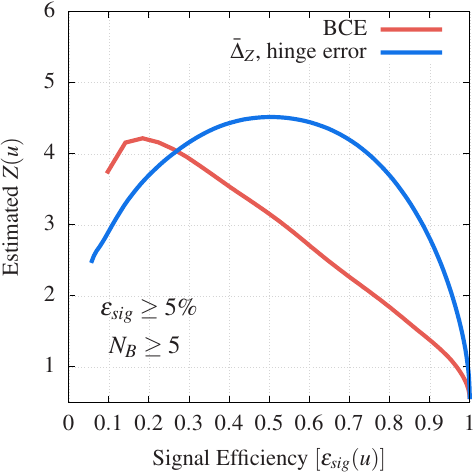}}\hfill
\subfigure[\hspace{-0.5cm}]{\includegraphics[width=0.31\textwidth]{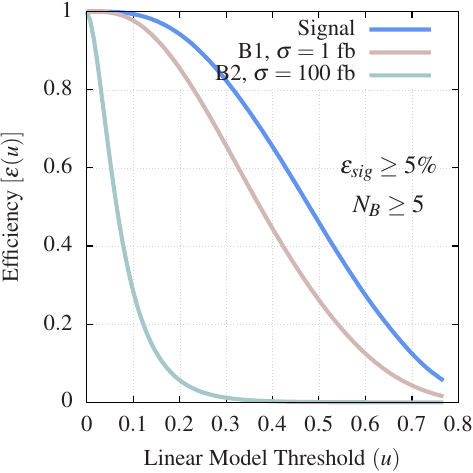}}\hfill
\caption{\label{fig:metrics}(a) The estimated $Z$ score for the entire range of the linear model threshold, $u$. (b) The distribution of the $Z$ score with signal efficiency, the fraction of signal events retained. Both quantities are functions of $u$. The model trained with $\bar{\Delta}_Z$ reaches the maximum $Z$ score for higher values of signal efficiency than that with the BCE loss. (c) Class efficiencies vs. $u$ when trained with the $\bar{\Delta}_Z$ loss with the hinge error. With increasing $u$, the larger background is eliminated first. For very high $u$ the drop in the subdominant background is less steeper than the signal, leading to the drop in $Z(u)$ in (a) for higher thresholds.}  
\end{figure*}

We analyse the loss function with a simple toy problem which can be easily mapped to the problem of event classification at the LHC. Our goal is to separate the signal ($s$) from background events using a linear classifier in the presence of multiple (say, two, $b1$ and $b2$) dominant background processes, as is usually the case. The datasets are modelled as normal distributions in two features, $x_1$ and $x_2$ which can be thought of as the kinematic features of the actual events. We generated roughly $50000$ points for each process and the optimisation was done in batches using \texttt{RAdam} optimiser. 
We train the linear classifier using the BCE loss and $\bar\Delta_Z$ with the hinge error for the following two test cases: 
\begin{itemize}
\centering
    \item[Case 1:] $\sigma_{b1} = 1$ fb, $\sigma_{b2} = 100$ fb; $\sigma_{s} = 0.1$ fb.
    \item[Case 2:] $\sigma_{b1} = 100$ fb, $\sigma_{b2} = 1$ fb; $\sigma_{s} = 0.1$ fb.
\end{itemize}
with $\mathcal L=3000$ fb$^{-1}$. We show the data distributions in Fig.~\ref{fig:Loss_DecBound}. Since $\bar\Delta_Z$ has the event rate (the true probabilities) information, we expect the decision boundaries to be different for the two test cases---eliminating more events from the larger background will give better significance scores, which $\bar\Delta_Z$ is designed to optimise. Fig.~\ref{fig:Loss_DecBound} confirms this. 

\paragraph{Performance:} To compare the performance of the BCE- and $\bar \Delta_Z$-trained models, we show the results of some scans in Case 1 in Fig.~\ref{fig:metrics}: the estimated $Z$ score value for different model thresholds ($u$, varying which essentially translates the decision boundaries in Fig.~\ref{fig:Loss_DecBound} along the axes) and the variation of the estimated $Z$ score with the signal efficiency ($\varepsilon(u)$, the fraction of signal events retained for the threshold $u$) against the $Z$ score obtained. For the scan, we demand $\varepsilon(u)\geq 0.05$ and the number of background events beyond the threshold to be at least $5$. Similar plots are obtained for Case 2 also. From the figure, we see that $\bar\Delta_z$ maximises the $Z$ score for a higher signal efficiency than the BCE, i.e., where the estimated $Z$ score peaks, the model retains more signal events than the BCE-trained model. (For the $\bar\Delta_z$ model, the estimated $Z$ score drops for high values of $u$ because there, for the datasets we consider, the major background is almost eliminated and further shifting the decision boundary reduces the minor background slower than the signal).

\section{Conclusions and Outlook}
In this article, we showed how a loss function for directly optimising the signal significance can be constructed. We obtained a surrogate for the median $Z$ score, proved that it is a submodular set function and derived a loss function that can be used to train a multivariate model in batches using the Lov\'asz extension. We showed that models trained with such a loss can cut the heavy background(s) more than the ones trained on the BCE loss while retaining more signal events (and thus showing the promise of enhancing experimental sensitivities).

\paragraph{Limitations and scope:}
While our results are promising, further tests are needed to fully characterise and understand the benefits and limitations of $\bar \Delta_Z$.  Here, our choice of using a linear classifier on simple datasets was motivated by its simplicity and interpretability. However, for realistic characterisations, one has to look beyond the linear classifier (e.g., use a deep neural network) and consider a range of benchmark (new-physics) scenarios with different kinematics (features). For example, there could be multiple (more than two) major backgrounds with highly overlapping features or the signal size could be much smaller than the backgrounds (more than what we considered, as is the case in some heavy particle searches). 

Finally, we note that while it is possible to introduce rate-dependent weights directly in the BCE loss, tuning them is an empirical task. The weights that yield the best performance need not be simply the rates of the processes. In contrast, $\bar{\Delta}_Z$ presents a natural way to include the rates (cross sections) as it is derived from the significance score used in collider searches.

\section{Code availability}
A \textsc{Python} implementation for $\bar{\Delta}_Z$ is available at \href{https://github.com/Jai2500/Z-Score-Loss}{https://github.com/Jai2500/Z-Score-Loss}.

\section{Acknowledgments}
J. B. would like to thank Arpan Dasgupta for valuable discussions on submodular functions. 

\bibliographystyle{plain} 
\bibliography{ref_v1}

\begin{thebibliography}{1}

\bibitem{DBLP:journals/corr/BermanB17}
Maxim Berman and Matthew~B. Blaschko.
\newblock Optimization of the jaccard index for image segmentation with the lov{\'{a}}sz hinge.
\newblock {\em CoRR}, abs/1705.08790, 2017.

\bibitem{Cowan:2010js}
Glen Cowan, Kyle Cranmer, Eilam Gross, and Ofer Vitells.
\newblock {Asymptotic formulae for likelihood-based tests of new physics}.
\newblock {\em Eur. Phys. J. C}, 71:1554, 2011.
\newblock [Erratum: Eur.Phys.J.C 73, 2501 (2013)].

\bibitem{lin2017focal}
Tsung-Yi Lin, Priya Goyal, Ross Girshick, Kaiming He, and Piotr Doll{\'a}r.
\newblock Focal loss for dense object detection.
\newblock In {\em Proceedings of the IEEE international conference on computer vision}, pages 2980--2988, 2017.

\bibitem{lovaszHinge}
Jiaqian Yu and Matthew Blaschko.
\newblock The lovász hinge: A novel convex surrogate for submodular losses, 2015.

\end{thebibliography}

\newpage

\appendix

\section{Proof of sub-modularity of the $\mathbf{Z}$ score}\label{App:Z_submodular_proof}
For the proof, we take the scenario with a single signal process ($|S| = 1$, $\mathbf{n} = n_1 = n$) and a single background process ($|P| = 1$, $\mathbf{p} = p_1 = p$) to simplify the expressions. But the result can be easily extended to incorporate multiple signal and background processes due to the linearity of additional signal processes and background processes. We will also drop the luminosity term as that will not affect the core derivation. 

For the proof, let us assume that we have two sets of misclassifications $C$ $(n_C, p_C)$ and $D$ $(n_D, p_D)$, such that $D \subseteq C$, i.e., 
\begin{equation}
    D \subseteq C, \quad n_D \leq n_C, \quad p_{D} \leq p_{C} 
\end{equation}
The total number of events remains the same between $C$ and $D$, i.e., $v_S$ for signal and $v_B$ for background, and only the misclassifications on the total set change.

To establish the proof, we need to show that the diminishing return property of submodularity holds under the addition of a new element $i \notin C$. 

\noindent\textbf{Case I: Adding false negatives $i \notin C$}

We prove that $\Delta_Z$ is submodular under the addition of false negatives:
\begin{align}
    \Delta_Z(C \cup \{i\}) &= \Delta_Z(n_C + 1, p_C) \\
    &= \frac{\sigma_S}{\sqrt{\epsilon}} - \frac{\frac{v_S - n_C - 1}{v_S} \sigma_S}{\sqrt{ \epsilon + \frac{p_C}{v_B} \sigma_B}} \\ 
    &= \left( \frac{\sigma_S}{\sqrt{\epsilon}} - \frac{\frac{v_S - n_C}{v_S} \sigma_S}{\sqrt{ \epsilon + \frac{p_C}{v_B} \sigma_B}} \right) + \frac{\frac{1}{v_S} \sigma_S}{\sqrt{ \epsilon + \frac{p_C}{v_B} \sigma_B}} \\
    &= \Delta_Z(C) + \frac{\frac{1}{v_S} \sigma_S}{\sqrt{ \epsilon + \frac{p_C}{v_B} \sigma_B}} \\
    \Delta_Z(C \cup \{i\}) - \Delta_Z(C) &= \frac{\frac{1}{v_S} \sigma_S}{\sqrt{ \epsilon + \frac{p_C}{v_B} \sigma_B}} \label{eq:LZproof-fn} 
\end{align}
Now since $D \subseteq C$, i.e., $p_D \leq p_C$, we see from Eq.~\eqref{eq:LZproof-fn}, 
\begin{equation}
\label{eq:lZproof-fn-proved}
    \Delta_Z(C \cup \{i\}) - \Delta_Z(C) \leq \Delta_Z(D \cup \{i\}) - \Delta_Z(D), \quad i \text{ is a false negative} 
\end{equation}

\noindent\textbf{Case II: Adding a false positive $i \notin C$}

We prove that $\Delta_Z$ is submodular under the addition of false positives:
\begin{align}
    \Delta_Z(C \cup \{i\}) &= \Delta_Z(n_C, p_C + 1) \\ 
    &= \frac{\sigma_S}{\sqrt{\epsilon}} - \frac{\frac{v_S - n_C}{v_S} \sigma_S}{\sqrt{ \epsilon + \frac{p_C}{v_B} \sigma_B + \frac{1}{v_B} \sigma_B}} \\ 
\end{align}
Now we have,
\begin{align}
    \Delta_Z(C \cup \{i\}) - \Delta_Z(C) &= \frac{\frac{v_S - n_C}{v_S} \sigma_S}{\sqrt{ \epsilon + \frac{p_C}{v_B} \sigma_B}} - \frac{\frac{v_S - n_C}{v_S} \sigma_S}{\sqrt{ \epsilon + \frac{p_C}{v_B} \sigma_B + \frac{1}{v_B} \sigma_B}}  \\
    &= \underbrace{\left( \frac{v_S - n_C}{v_S} \sigma_S \right)}_{T_1} \underbrace{\left[ \frac{1}{\sqrt{ \epsilon + \frac{p_C}{v_B} \sigma_B}} - \frac{1}{\sqrt{ \epsilon + \frac{p_C}{v_B} \sigma_B + \frac{1}{v_B} \sigma_B}} \right]}_{T_2}
\end{align}

Therefore it decomposes into a product of two terms. If we show that independently both of these terms are independently smaller for $C$ than for $D$, we will have our result.

First consider $T_1$, we have
\begin{align}
    n_C &\geq n_D \\
    = -n_C &\leq -n_D \\
    = \frac{v_S - n_C}{v_S} &\leq \frac{v_S - n_D}{v_S}
\end{align}

\noindent Therefore, $T_1$ is indeed larger for $D$ compared to $C$.

\noindent In order to check for term $T_2$, we first simplify the expression and write $H_C = \epsilon + \frac{p_C}{v_B} \sigma_B$, ($H_C \geq H_D$). Now we can write term two as:
\begin{equation}
    \frac{1}{\sqrt{H_C}} - \frac{1}{\sqrt{H_C + \frac{1}{v_B} \sigma_B}} \label{eq:LZproof-simplified}
\end{equation}
We move to a continuous relaxation of the term such that:
\begin{align}
    f(x) &= \frac{1}{\sqrt{x}} - \frac{1}{\sqrt{x + \frac{1}{v_B} \sigma_B}}  \label{eq:LZproof-contrelax} \\ 
    f(H_C) &= \frac{1}{\sqrt{H_C}} - \frac{1}{\sqrt{H_C + \frac{1}{v_B} \sigma_B}}
\end{align}
which is the same as Eq.~\eqref{eq:LZproof-simplified}. Now differentiating Eq.~\eqref{eq:LZproof-contrelax} with respect to $x$, we get:
\begin{equation}
    \frac{\text{d}}{\text{d}x} f = \frac{1}{2} \left( \frac{1}{\left({x + \frac{1}{v_B} \sigma_B}\right)^\frac{3}{2}} - \frac{1}{\left(x\right)^\frac{3}{2}}  \right)
\end{equation}
which will always be less than zero for $x > 0$. Thus since $\frac{\text{d}}{\text{d}x} f < 0$, we have that $T_2$ will be greater for $D$ compared to $C$.

Now since both $T_1$ and $T_2$ is greater for $D$ compared to $C$, we have
\begin{equation}
    \Delta_Z(C \cup \{i\}) - \Delta_Z(C) \leq \Delta_Z(D \cup \{i\}) - \Delta_Z(D), \quad i \text{ is a false positive} \label{eq:LZproof-fp-proved}
\end{equation}

Therefore from \textbf{Case I} and \textbf{Case II}, we have shown that $\Delta_Z$ is submodular for all the possible cases and therefore is submodular for the set of misclassifications $(\mathbf{n}, \mathbf{p})$. \hfill

\section{Error Functions}
\label{app:error_fns}
We require a loss function to handle any vector of errors $\mathbf{m} \in \mathbb{R}^p_+$ since we are working with continuous predictions, not only to discrete vectors of misclassifications in $\{0, 1\}^p$. We consider four cases for defining the vector of errors $\mathbf{m}$ to construct the surrogate losses using the Lovasz extension. 
\begin{enumerate}
    \item \textbf{Hinge (Max Margin) Loss:} Following Ref.~\cite{lovaszHinge}, we implement a hinge loss to compute the error in the prediction. The labels are considered signed ($y_i \in \{-1, 1\}$). The model outputs a score $F_i(x)$ for each sample $x$. The error is given by the hinge loss, 
    \begin{equation}
        m_i = \max(1 - F_i(x)y_i, 0), \qquad y_i \in \{-1, 1\}.
    \end{equation}
    \item \textbf{Sigmoid Error}: Similar to Ref.~\cite{DBLP:journals/corr/BermanB17}, we also consider the sigmoid error. The model outputs a probability $F_i(x)$ for the sample $x$ to be in the signal class. The error is given by 
    \begin{equation}
        m_i = \begin{cases}
            1 - F_i(x), & \text{if } y_i = 1, \\
            F_i(x), & \text{otherwise.}
        \end{cases}
    \end{equation} 
    \item \textbf{Cross Entropy Error}: We also experiment with the BCE loss to measure the error $m_i$. This is similar to taking the logarithm of the error calculated in the Sigmoid Error. One could also interpret this as a form of weighted cross entropy where the weights are calculated based on the specific composition of the batch of events and misclassifications on that batch.
    \item \textbf{Focal Loss Error}: We also consider Focal Loss ~\cite{lin2017focal} as the measure for the error $m_i$. This loss function drives the network to focus on hard misclassified events. 
\end{enumerate}

\begin{figure*}
\centering
    \subfigure[Hinge Error]{
        \label{fig:loss-scape-hinge}
        \includegraphics[width=0.48\linewidth]{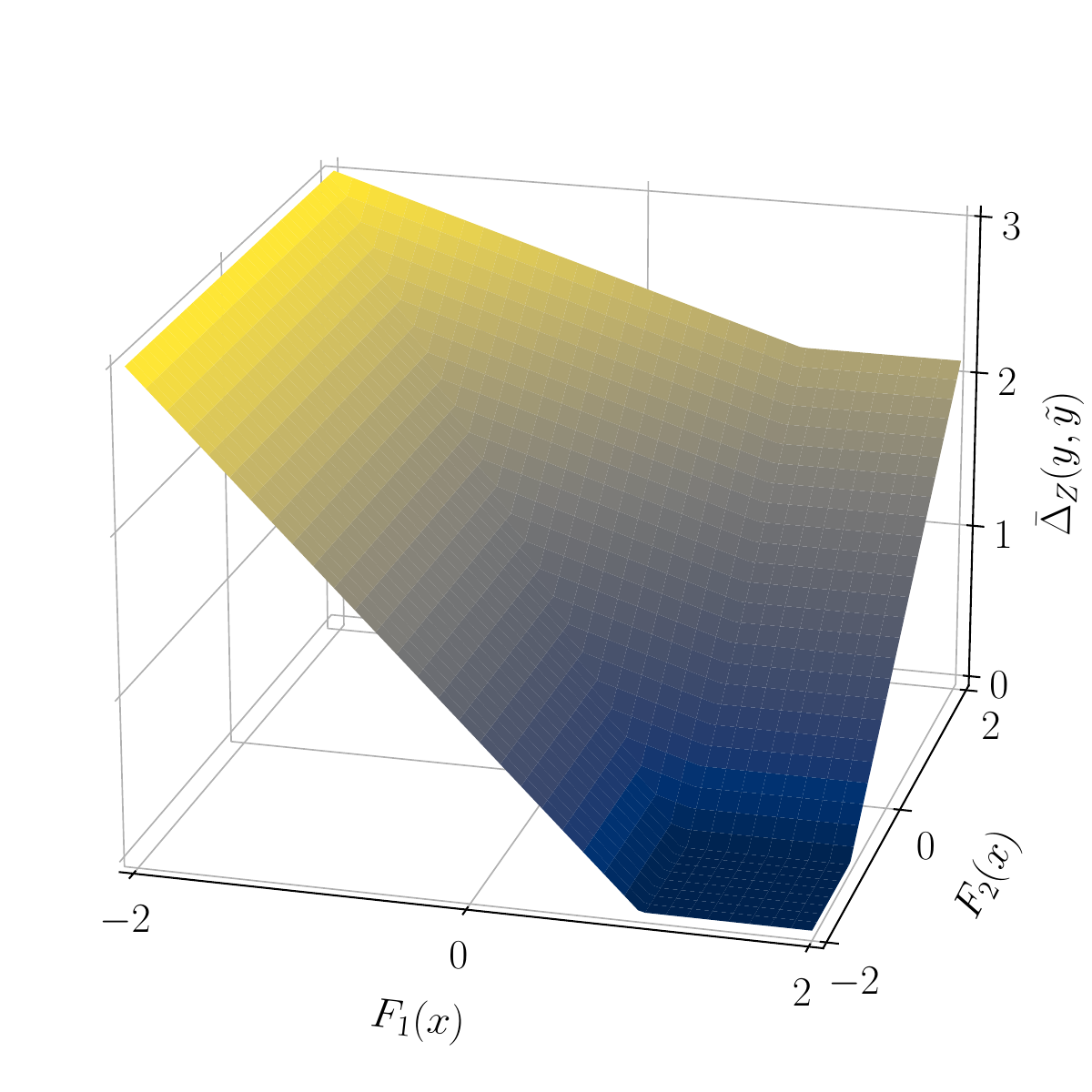}
    }\hfill
    \subfigure[Sigmoid Error]{
        \label{fig:loss-scape-sigmoid}
        \includegraphics[width=0.48\textwidth]{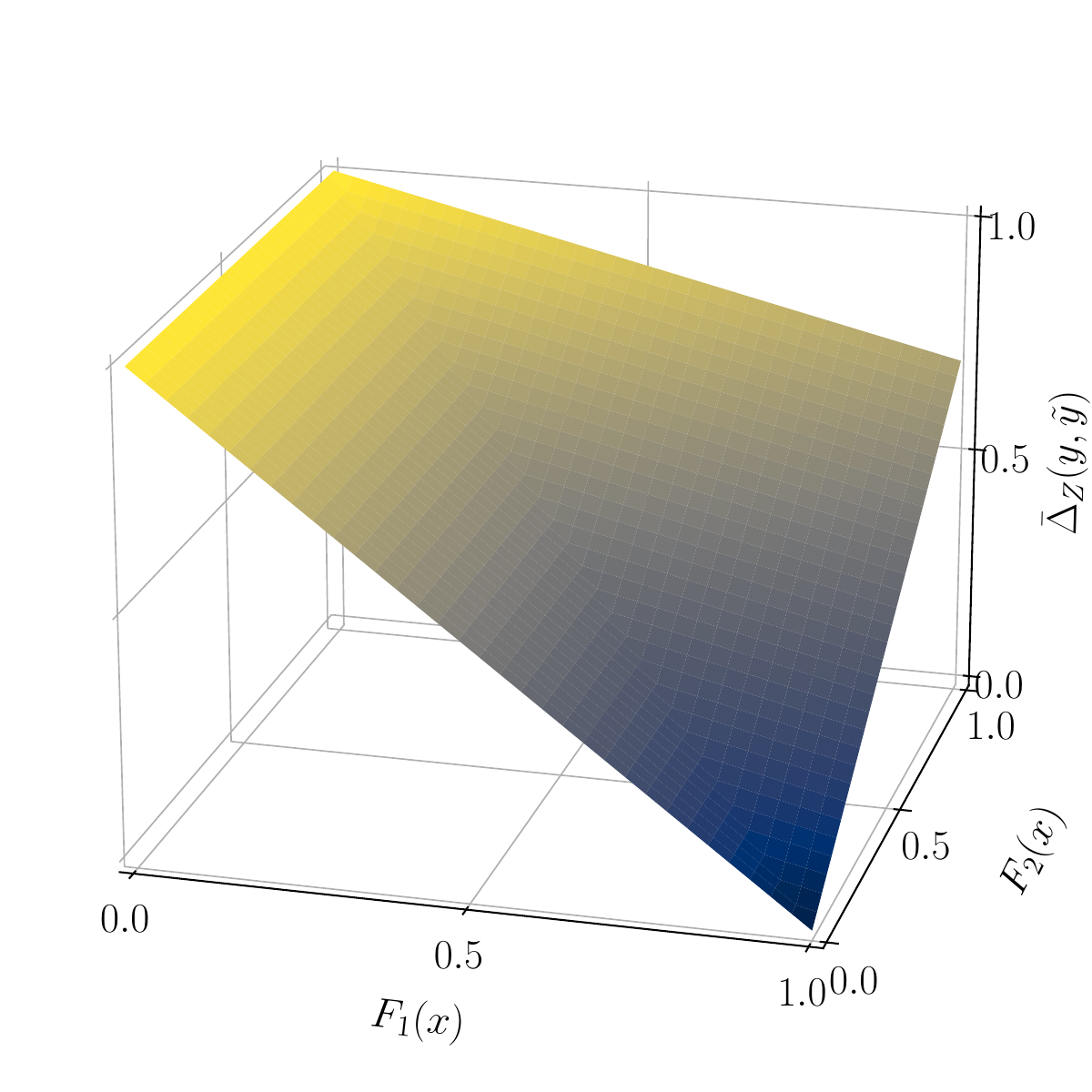}
    }\hfill
    \subfigure[CE Error]{
        \label{fig:loss-scape-ce}
        \includegraphics[width=0.48\linewidth]{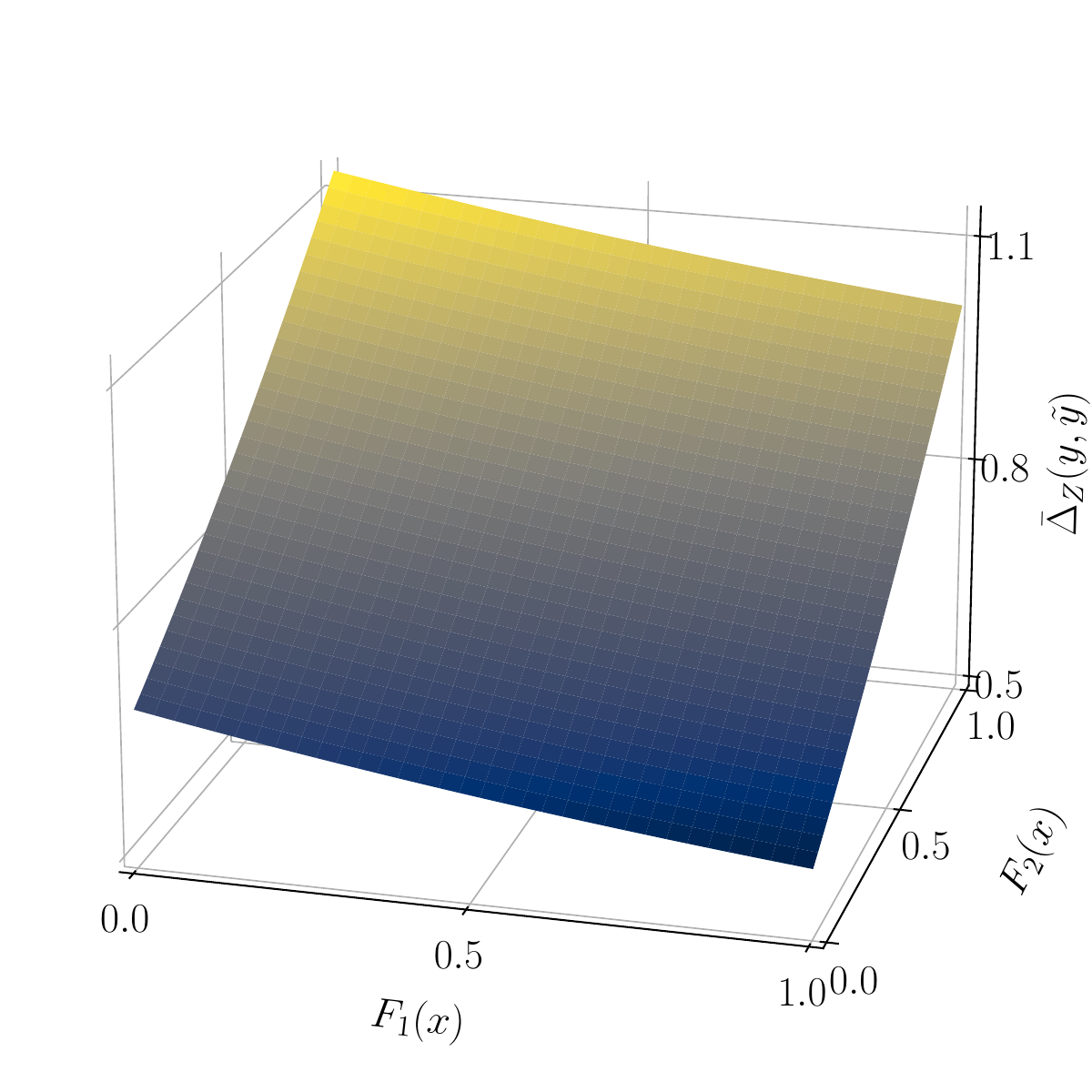}
    }
    \subfigure[Focal Error]{
        \label{fig:loss-scape-focal}
        \includegraphics[width=0.48\textwidth]{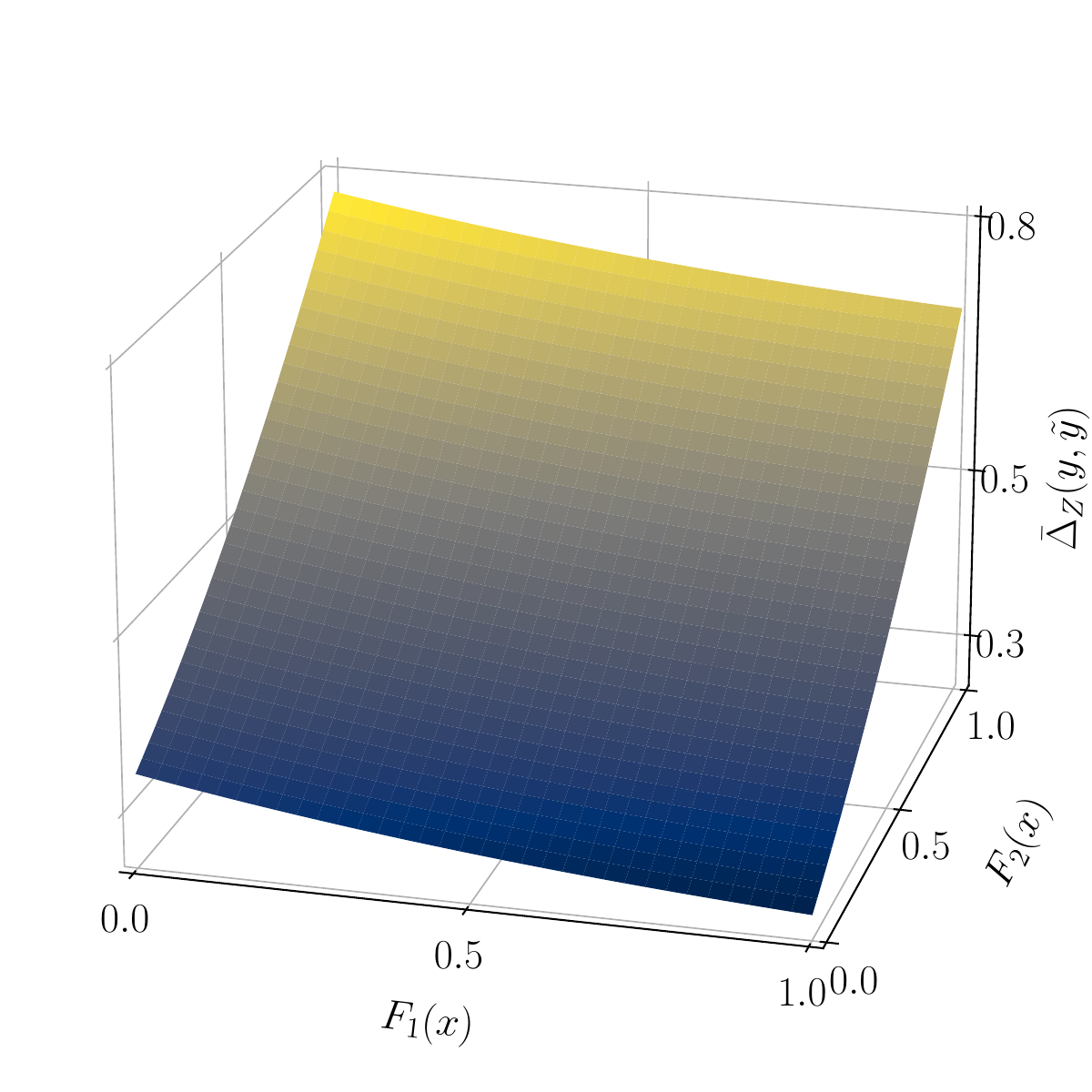}
    }
\caption[Loss landscapes for the four error measures $\mathbf{m}$]{\label{fig:LZViz}Loss landscapes for the four error measures $\mathbf{m}$ in Sec.\S~\ref{sec:LZerr}. The $Z$ score loss is plotted with ground truth, $GT = [1, 0]$, $\sigma = [1, 10]$. The $x, y$ axes denote the classifier output ($F_1(x), F_2(x)$). For the Hinge Error, the GT labels are converted to their signed equivalent.}
\end{figure*}

Our formulation results in a convex loss with a global minimum, which we evaluate for a simple case for all the error metrics as visualized in Fig.~\ref{fig:LZViz}.

\section{ROC Curves for Case 1.}

\begin{figure}[th]
\centering
\subfigure[\hspace{-0.75cm}]{\includegraphics[width=0.48\textwidth]{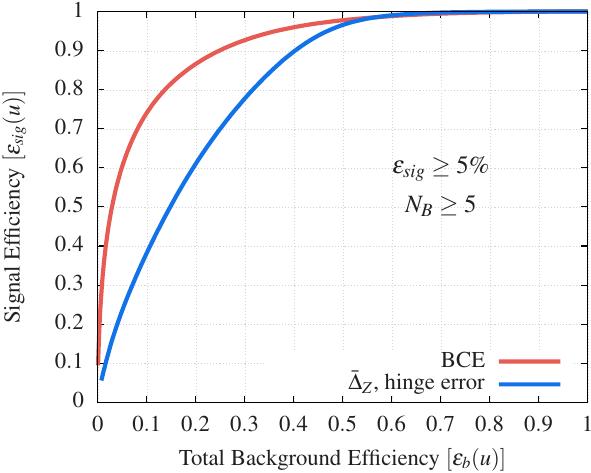}}\hfill
\subfigure[\hspace{-0.75cm}]{\includegraphics[width=0.48\textwidth]{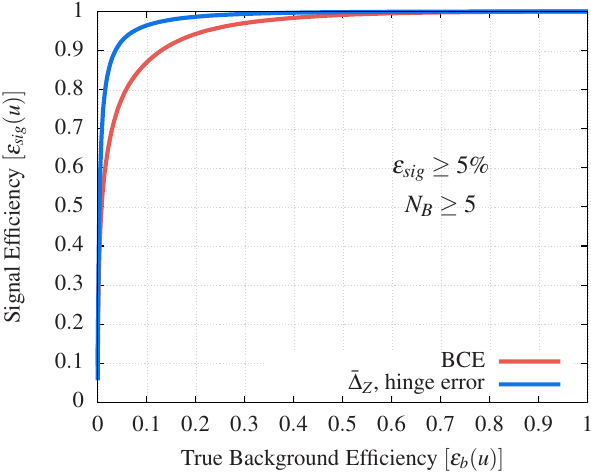}}

\caption{(a) ROC Curve for dataset (total) background efficiency vs signal efficiency. (b) ROC Curve for true background efficiency vs signal efficiency. The true background efficiency differs from the total background efficiency in that it accounts for the cross sections of the background processes. We observe that our loss performs better at removing background at a higher signal efficiency.} \label{fig:roc_curve}
\end{figure}

We plot the ROC curves for experiments for Case 1 in Fig.~\ref{fig:roc_curve}. Case 2 gives similar results. Let $N_{B1}, N_{B2}$ represent the total number of BG1 and BG2 events generated in the dataset. Let $n_{B1}, n_{B2}$ represent the number of BG1 and BG2 events remaining after the threshold respectively. Let $\sigma_{B1}, \sigma_{B2}$ represent the cross sections of process BG1 and BG2 respectively. The total background efficiency is given by,
\begin{equation*}
    \frac{n_{B1} + n_{B2}}{N_{B1} + N_{B2}},
\end{equation*}
and the true background efficiency is given by,
\begin{equation*}
    \frac{\left(\frac{n_{B1}}{N_{B1}}\right) \sigma_{B1} + \left(\frac{n_{B2}}{N_{B2}}\right) \sigma_{B2}}{\sigma_{B1} + \sigma_{B2}}.
\end{equation*}

\end{document}